# Decision Support Systems – Technical Prerequisites and Military Requirements

**Dr. Andreas Tolk**
Industrieanlagen-Betriebsgesellschaft mbH (IABG)
Einsteinstr. 20
D-85521 Ottobrunn, Germany
+49-89-6088 2381
tolk@iabg.de

**Major Dietmar Kunde**
Heeresamt (German Army Office) HA V (3)
Bruehler Str. 300
D-50968 Koeln, Germany
+49-221-9371 3214
DietmarKunde@bwb.org

## Abstract

*Decision Support Systems in the sense of online alternative course of action (ACOA) development and analysis as well as tools having the potential to be used for online Development of Doctrine and Tactics Techniques, and Procedures (DTTP) for support to operations*
- *make it possible to evaluate the command and control processes and the performance capabilities of the friendly and enemy forces and other decision relevant factors*
- *support the military commander (brigade and higher) and his staff in their headquarter by increasing their ability to identify own opportunities*
- *support all phases of the command and control process*
- *use computer based, automatic and closed models, that can be adapted to the current situation.*

*This paper presents the results of studies conducted in Germany on behalf of the German Ministry of Defense. Main tasks were to contribute to the conceptual basis for decision support systems within the German Army, to evaluate the influence of decision support systems on the command and control process and to consider international work as well.*

*The technical and operational requirements are described in detail that have to be met in order to support the warfighter within his command and control process. The planned and used means as well of applied Operations Research range from simple optimization algorithms to complex simulation federations comprising different systems.*

## 1 Introduction

This paper is dealing with Decision Support Systems helping the warfighter[1] to gain and maintain information superiority in order to achieve command superiority. The authors are convinced, that Decision Support Systems in the sense of online ACOA development and analyses as well as possible development of DTTP have to become an integral part of future C4I systems. They comprise several means of applied military Operations Research being used to translate the awareness of the battle space, i.e., perceiving what is occurring, understanding what is taking place and communicate the findings quickly, surely, accurately, and in an understandable and usable form to the combat forces. They are going to be used by the commander and his staff to support them in all situations of peacetime, crisis and war. They also help the commander in harmonizing his efforts with other – possibly being non-military – organizations, coordinate the overall allied workflow, and facilitate the access to all necessary forms of available information, including open sources.

---
[1] When talking about the warfighter, we are addressing the military user within an operational context without limitations, neither to special command levels nor to type of operation.

Among possible means of Operations Research, simulation systems play a special role due to their wide acceptance in different application domains in the armed forces. There are four main fields that have to be considered, especially when simulation systems are used as a sort of add-on to a C4I system to support the warfighter directly within – e.g. – a brigade or division headquarter:
- All processes of command, control, communications, intelligence, reconnaissance, attrition, movement, etc. relevant to the problem to be solved must be modeled adequately.
- Command agents and computer generated forces (CGF) have to be used for automatic order generation and intelligent behavior of simulated entities.
- The initial state of the simulation must be generated automatically out of the data available from the C4I system.
- Adequate and validated data must be available for the simulation system.

This paper describes the requirements of all four fields and gives some tentative recommendations on how to be able meet the respective challenges. In addition, the need for harmonization of the data and object models of the C4I world and the M&S world has been formulated in several publications. The introduction of a central data model to be used within all applications and databases is – from the practical point of view – hardly feasible. An appropriate alternative is to focus on data interchange and to provide a shared data model accompanied by data mediation/translation and system migration procedures. As a shared data model represents a common and continuously evolving understanding of information, it must be flexible, open, and extensible. This challenge also has to be faced and solved when integrating several independently developed applications to give support to operations.

Last but not least the requirements for support to operations cannot be derived from technical aspects only. They have to be real user needs being mainly derived from the commander's and his staff's needs for support within real operations. However, a good way seems to be the use of CAX environments to find out where the real needs of the users are, what the several simulation systems are able to contribute to the user and what has to be done to eliminate the lack the systems still have to be used as decision support systems.

This paper summarizes the findings of the work having been done in Germany over the last approximately five years. Additional information can be found in papers with more specific topics, e.g., [Tolk, 1999a], [Krusche and Tolk, 1999], [Tolk and Schiefenbusch, 1999] and [Tolk, 2000] as well as in respective study reports available to the Federal Armed Forces.

In addition, a lot of national and international studies were initiated over the last months dealing with the possibility of using modeling and simulation for support to operations, among which the NATO Studies, Analyses, and Simulation Panel Activity should be mentioned, trying to deliver a NATO Code of Best Practice (COBP) for C2 Assessment in Operations Other than War. Another example on NATO level is the SAMOC software containing simulation software for training as well as for support to operations. As an actual US effort, the Version X of the Army Battle Command System (ABCS) has to be mentioned. The ABCS explicitly requires support to operations by ACOA evaluation using integrated simulation components.

To summarize this, it can be said that the need for integrated Operations Research support meanwhile is an issue for C4I systems' development.

## 2      General Issues of Decision Support Systems

There are at least two views when dealing with Decision Support Systems:

- The warfighter point of view asks, how such tools can help him to reach his objectives and fulfil his tasks better, faster, and binding less resources.
- The technical view focuses on information technology and applied systems science or operations research issues to be solved.

This paper deals with the military requirements and the technical aspects as well. The issue, which aspect is the driving one, is beyond the scope of this paper. This section gives a current definition of Decision Support Systems and prepares the floor for the technical and military aspects. The following statements do not claim completeness at all.

### *2.1      Definition of Decision Support Systems*

The support of the military commander and his staff in the decision making process has to be the main objective of respective systems in order to increase the overall efficiency of the integrated system of Command and Control – Reconnaissance – Effects. Decision Support Systems contribute essentially to the Command and Control Superiority.[2]

The integration of Operations Research methods into the Command and Control Process is a prerequisite to enable more rapid decision cycles and to improve qualitatively the basis of the decisions to be made. To support the Command and Control process safely and quickly it's necessary to have comprehensive knowledge of the own situation and a sufficient good estimate of the enemy situation. Decision cycles must take place quickly and provide high "quality" decisions, in order to gain and keep the initiative in combat as well as in operations other than war (OOTW).To this end, existing decision support systems are add-ons to existing C4I-systems, but in the medium and long term they must become integral parts. Currently decision support systems in the sense of ACOA are defined as follows:

> **Decision Support Systems are applied OR methods for the support of the military commander and his staff. They support all phases of the Command and Control process by providing and assessing information obtained from the respective C4IS in war and peace-time.**

---

[2] It should be pointed out, that in the scope of this paper the main objective of the use of decision support systems is to gain information superiority in order to achieve command and control superiority. Information superiority per se isn't seen as a value. Only in the operational context of being able to achieve better results faster, information superiority becomes valuable.

During OOTW, in addition to the well known procedures of military operations, the coordination with non-military organizations, the harmonization of the workflow, the controlled dissemination of information following the "need-to-know" principle (incl. national commanders, media, etc.) have to be assured and, therefore, supported by the operational C4I systems as good as possible. Again, OR methods can be used to do so.

Therefore, the overall objective is to enable timely and effective decisions in every situation by supporting the commander and his staff with the information they need in the form they need it. This leads to the first general issue to be handled, the information overflow on the battlefield.

## 2.2 Classes of Decisions

The German study [IABG, 1999] as well as the NATO COBP [NATO, 1999] are distinguishing between several classes of decisions that can be supported by respective means of Operations Research. Already the classical decision theory structures the decision for structured approaches and solutions. We want to introduce three different approaches.

First way to structure decisions is to look at their structure. A decision may be complete, that means comprising all elements and sub-decisions belonging to the actual decision to be made, e.g., an operations plan, or it may be part of a comprising decision set, e.g., the decision how to engage the artillery within an operation. Some literature references this distinction also as open and closed decisions. It had to be said, however, that it is not possible always really to decide whether we are looking at an open sub-decision or whether the operation is a closed complete one. An operation plan, e.g., seems to be closed, however, he is as a rule part of a comprising campaign, hence he is open. However, looking at where the most decision support is needed is in the domain of complete decisions, whereas looking at where most means of Operations Research can be applied are clear defined sub-decisions, e.g., route optimization, etc.

Second way to structure the classes is looking at the decision trigger type, i.e., which cause or reason results in the launch of a respective decision process. Decisions may be event triggered and therefore unique or single decisions, handling within just a given situation and being hardly "reusable", or they may be routine decisions, e.g., route optimization, ammunition storage. Again, the most needed support is in the domain of unique decisions whereas Operations Research offers help for the area of routine decisions.

Third way to structure classes of decisions is to look at the environment under which the decisions have to be made. Literature distinguish between decision under certainty (everything is known to the decision maker), decisions under uncertainty risk (there are several alternatives that may occur, however, the decision maker just knows the likelihood of everyone), or decisions under risk uncertainty (where nothing may be known). As before, there is a good amount of help available for decisions under certainty and even under uncertaintyrisk, but most of the help is needed in the domain of decision making under riskuncertainty. Respective recommended ways, procedures, and solutions can be found, e.g., in [NATO, 1995].

On behalf of the Research and Technology Organization (RTO) of NATO, the Studies, Analysis, and Simulation (SAS) Panel establish a research study group (RSG) on C2 assessment. The RSG developed a NATO Code of Best Practice (COBP) dealing with all aspects of modern command and control, not only in the scope of information superiority [NATO, 1999]. Thise NATO COBD, however, uses a more user or military operation oriented way of structuring decisions. These distinctions have also found their way into the German considerations. The COBP mentions simple decisions, automatable decisions, and complex decisions, that are defined as follows:

- Simple Decisions mean "to know is to decide". There are no real alternatives to take into account. They are routine decisions under certainty and can in principle be taken over by machines or computers.
- Automatable Decisions mean "to know is to decide, but knowing is not yet possible". These are typical contingency decisions or routine decisions under uncertainty. Machines or computers can at least help the decision maker not to miss the right decision points, to take the effects of the different alternatives into account, etc.
- Complex Decisions are the domain of the decision maker. They cannot be taken over by machines, although decision support is possible.

In this sense, decision support systems can take over the simple and automatable decisions and, in addition, can support the military decision maker within the domain of complex decisions.

To summarize this section: Decision Support Systems as applied means of Operations Research can support the military decision maker by taking away simple and pre-planned decisions and supporting him within his real domain: non-automatable, complex decisions.

### 2.3     *Information Overflow on the Battlefield*

Although in former times the main problem was to get the necessary information, nowadays it is more the problem to find the right and appropriate piece of information within the huge amount of information being available from the different sources.

Not only the number of sensors has increased, also the number of information interchange sources as well as interoperable systems is much bigger then before. New sources – including the open sources of the internet – challenging the warfighter as well as his technical support crew with new requirements. The NATO Virtual Command and Control Center (VCC) as well as the US Virtual Information Center (VIC) are among the first applications dealing with the problem of combining traditional military information sources as well as open sources, including first approaches to face the problems of information insurance, security issues, etc.

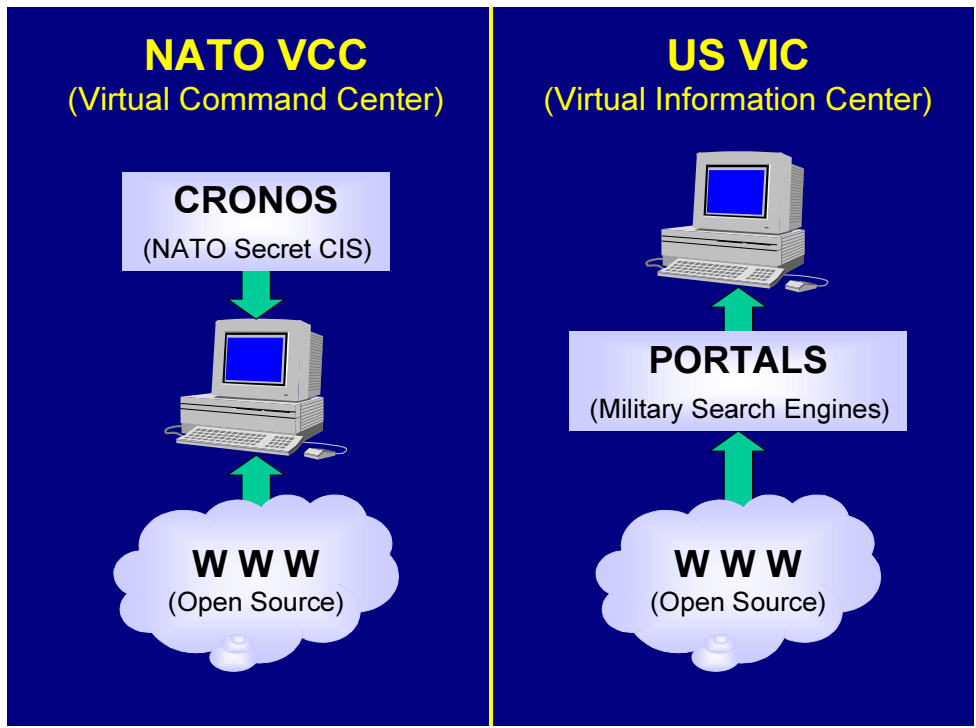

**Figure 1: VCC and VIC using Open Sources**

Thus, the military commander needs tools to support him in finding the necessary information and presenting them to him in the needed and desired form, i.e., graphically, in form of tables, or may be increasingly using multi media options in the future.

The NATO COBP [NATO, 1999] defines three main domains for Command and Control Systems to support the Commander
- Battlespace Visualization,
- Battle Management, and
- Decision Making.

The following figure depicts the different domains and respective functional categories that should be comprised within modern Command and Control Systems. It is based on the COBP functional categories of C4I systems introducing new categories for OOTW.

Overall functionality being captured within an information grid supports the warfighter in finding, presenting, coordinating, etc. the information. Based on this information, the commander can understand, what is going on and develop several alternative courses of action (ACOA). The C4I system should help him further in establishing his objectives, assess his former developed ACOAs, manage the uncertainty and finally, by setting respective criteria deciding and making a plan. This plan has to be approved and disseminated. Last but not least, the intelligence, surveillance and reconnaissance (ISR) aspects have to be taken into account using the appropriate priorities. The necessary functionality is quite well known from article five operations.

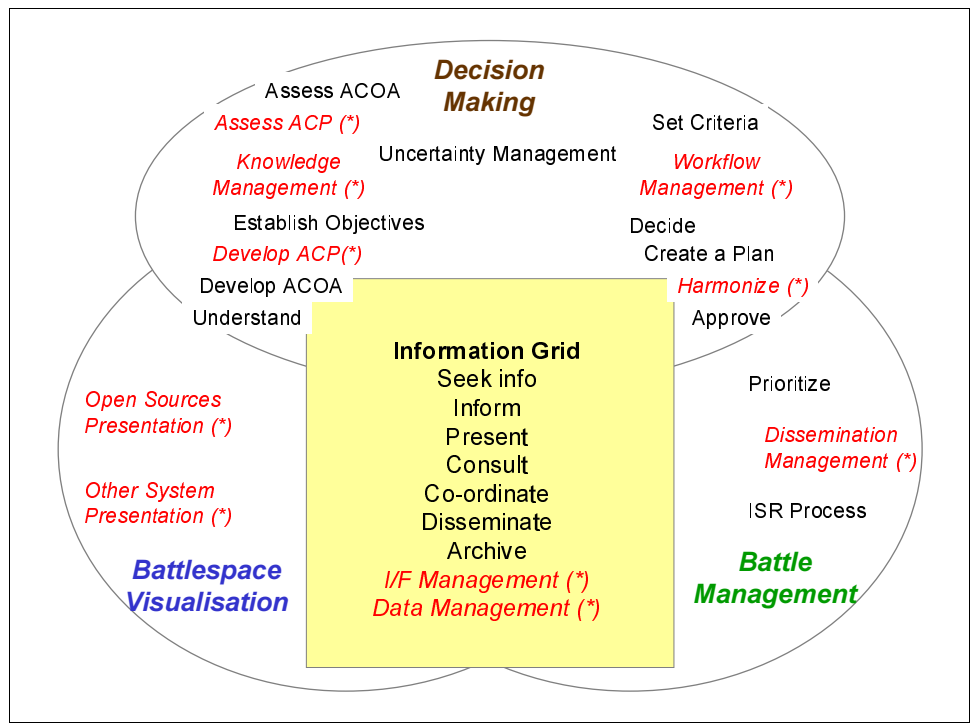

**Figure 2: Functional Categories of C4I systems**

The red italic function groups (followed by the asterisk) are new to highlight the application domain of OOTW. As before, overall functionality is comprised in the information grid. As in joint and combined operations it is never sure, who your partners will be. It may be also non-military organizations. Thus, interface management is necessary to enable technical interoperability. In order to assure aligned data understanding on the semantic level, in addition tools for online ad hoc data management are necessary bridging the time until standardized data elements for data exchange have been agreed to.

In the domain of battlespace visualization, other systems as well as open sources, especially the WWW and media, have to be dealt with. It is necessary to be able to access this information and display it in convenient form to the commander and decision maker.

In the domain of decision making, the development of alternatives contingency plans (ACP), in addition to the alternative courses of action (ACOA), have to be developed and assessed. Recent studies in the US pointed out, that the advantages of the digitized battlefield only can be used when appropriate contingency plans exists to which the decision maker can switch in case of need. Furthermore, knowledge management – including actual rules of engagement, status of partners, allies, political constraints, etc. – is necessary. As very different organizations may have to work together, workflow management to align the procedures, decision points, person to involve, etc. is necessary also. In the end, harmonization of all decisions of all organizations is needed, including the consistency check before the background of the actual degree of freedom for the decision in the actual – political – environment.

In the domain of battle management, the dissemination of the information have to be assured. This is not only true for the different lines of communication to, e.g., the operational controlling commander, the national commander, the political instances, but also to the media.

Decision Support Systems therefore can be seen as the general term for functional categories comprising means of Operations Research to support the military commander by establishing the needed functionality to enable him reaching his operational aims.

*2.4      The Need for Harmonization*

Interoperability is not a new requirement. However, in the era of joint and combined operations this requirement gains a new quality becoming an unavoidable necessity. In order to be able to use Operation Research tools within C4I systems, they have to be able to interchange information. Therefore, they need a common understanding of the information, preferable by using the same syntax and semantics in form of a common shared data model [Hieb and Blalock, 1999; Timian *et al.*, 1999; Krusche and Tolk, 1999; Tolk, 1999b]. In order to reach this objective, not only technical standards are needed, but the respective procedures of C4I and M&S development and implementation have to be aligned also. In addition, ways of migration for legacy systems have to be developed.

The Study Group on Interoperability between C4I System and Simulation Systems (SG-C4I) of the Simulation Interoperability Standardization Organization (SISO) developed a framework to cope with this issues [Lacetera and Timian, 2000]. The following figure – introduced by Michael R. Hieb and Andreas Tolk – comprises the fields to by harmonized and coped with to come to shared solutions.

First thing to be done is the alignment of architectures, so that components of both worlds are able to talk to another. The next level comprises common data and object models as well as common tools and common standards. This will lead to reusable components. However, to be able to reach real interoperable shared solutions, the processes have to be aligned also (e.g., using the same tools and methodologies) including procedures to migrate legacy systems to this new common world. Thus, more or less a change in philosophy of looking at C4I and simulation systems may be needed, e.g., when looking at M&S in acquisition, requirement analyses, support to operations, and training. Maybe, on the long term there will be no longer the distinction between both worlds but new systems will comprises functionality of both worlds as federates.

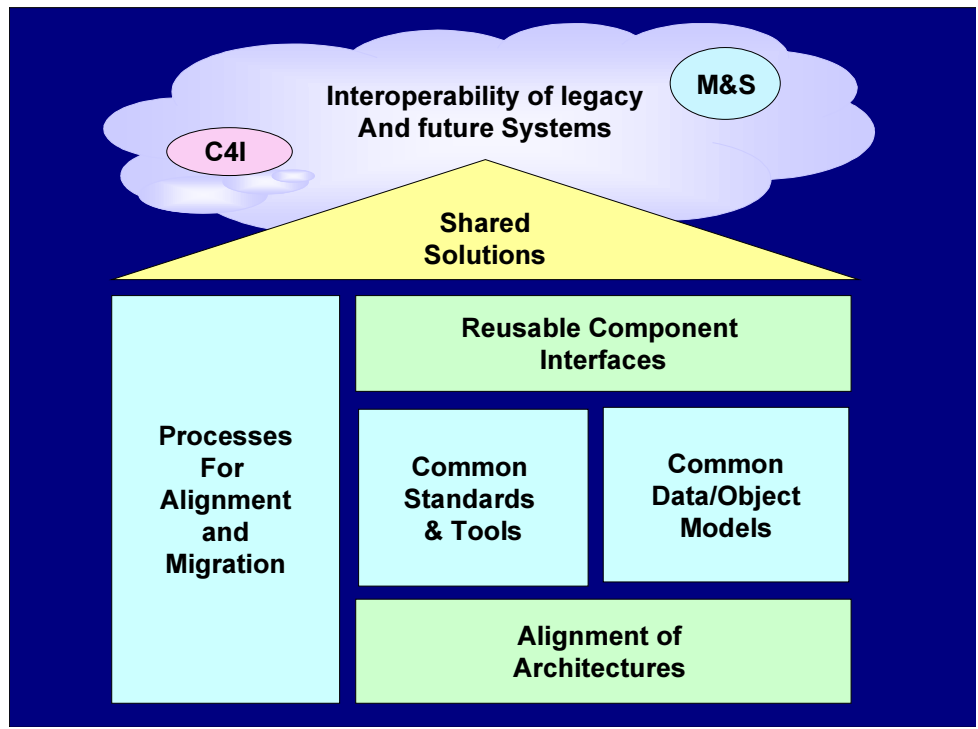

**Figure 3: Interoperability Issues of Shared Solutions**

## 3 Technical Issues of Decision Support Systems

What has been said in the beginning of the second chapter is true for the technical issues also. Among a lot of technical issues, three being of high relevance to the authors has been chosen to be highlighted within this paper. These are
- the need for a common understanding to be gained by common management processes,
- the possibility to use the result directly in an information integration layer, and
- technical requirements for simulation systems when being used for support to operations in the defined context.

The first two topics are necessities for the implementation of shared or federated solutions . Topic three shows new classes of requirements emerging from the domain of support to operations. These three topics will be covered in the following sections.

### 3.1 *Common Information and Data Management*

As pointed out in [Krusche and Tolk, 1999], generally each organization in the domain of defense depends on access to information in order to perform its mission. It must create and maintain certain information that is essential to its assigned tasks. Some of this information is private, of no interest to any other organization. Most organizations, however, produce information that must be shared with others, e.g., operation plans, location and activity of a given unit, information on the logistics, etc. This information must be made available, in a controlled manner, to any authorized user who needs access to it. At present, almost every defense information infrastructure exists as a collection of heterogeneous, non-integrated systems. This is true for C4I systems as well as for simulation systems and other means of Operations Research, and – when trying to bring them together – the problem of interconnections even increases. This is due to the fact that each organization builds systems to meet its own information requirements, with little concern for satisfying the requirements of others, or of considering in advance the need for information exchange. If any information exchange takes place, however, as a rule this information exchange is based on *ad hoc* interfaces. The result is an extremely rigid information infrastructure that costs months and millions to be changed or extended, and, which cannot cope with the increasing demand for widely integrated data sharing between multiple mission-related applications and systems. Actual solutions cannot solve these problems, thus, new ways have to be found in the era of joint and combined operations.

The Shared Data Environment (SHADE) fully described in [DoD, 1996] is a strategy to promote C4I systems' interoperability through a global view on the data of the battlespace, which is made available, in a controlled manner, to any authorized user who needs access to it. The objective is to define a *global infosphere*, that supplies a fused, real-time, true representation of the battlespace, to allow for an integrated data sharing between multiple mission-related applications and systems. The SHADE's technical focus and priorities are driven by near term systems' integration, migration and interoperability requirements that are identified in the Defense Information Infrastructure (DII) Common Operational Environment (COE) context. The main conceptual features of the SHADE address data interoperability for federations of system components and systems in general, not restricted to C4I systems. Thus, the ideas of SHADE become applicable to C4I systems as well as to simulation systems and other means of Operations Research as well. In order to do so, standard data elements (SDE) are defined for information exchange. In order to be able to manage this SDEs, a common shared data model is needed comprising all SDEs and giving them a semantic context.

A data model being able to cope with the requirements for the common shared data model has to have the following qualities:
- It must capture the information requirements of a wide range of battlefield functional areas. A common shared data model is best characterized as a "to-be" model of the required battlefield information rather than a model that is constructed with direct reference to existing current needs for information exchange.
- For flexible integration of future information (exchange) requirements, the data model must be constructed in a way that future information elements simply extend the model while its existing structure remains unchanged.

At has been shown in several publications, e.g., [Krusche and Tolk, 1999, Tolk, 1999b], the ATCCIS Generic Hub [NATO, 1996] meets both requirements quite well, as it has been designed to meet exactly these requirements by data modeling experts of almost all nations in NATO during the last 10 years. A tool-oriented view is given in [Tolk, 2000].

As has been pointed out, the definition of standard data elements (SDE) required for information exchange, the coordination and control of their implementation and use within systems have to be central objectives of an overall data management organization. They may not longer be under the responsibility of system managers who's legal and understandable objective is to optimize their system and, logically, neglecting often the requirements of the superimposed federation of systems.

In general, data management is planning, organizing and managing of data by defining and using rules, methods, tools and respective resources to identify, clarify, define and standardize the meaning of data as of their relations. This results in validated standard data elements and relations, which are going to be represented and distributed as a common shared data model.

The overall objective to be reached by introducing a data management is, to coordinate and to control the numerous system projects technically and organizationally, in order to improve the integrity, quality, security and availability of standard data elements. Due to this objective, the following central tasks of the data management organization are proposed:
- Definition of standard data elements across system boundaries,
- Evolutionary development of a common shared data model as a reference representation for standard data elements,
- Representation of standard data elements through a common shared data model,
- Definition of rules and methods for
  - access, modification and distribution of standard data elements,
  - introduction of new information exchange requirements,
  - Coordination and Control of system projects using the standard data elements in order to assure their consistent use and interpretation within different applications and systems.

To summarize, in order to reach the objective of a common shared data model comprising the standard data elements of the application domain, a respective common data management organization comprising central elements is essential.

### 3.2  *A General Information Integration Layers for Defense Applications*

Assuming that such a data management agency or organization uses the appropriate tools to do their work (see, e.g., [Krusche and Tolk, 1999; Tolk, 1999b; Tolk, 2000]), these results can be directly used to implement a general information integration layer. To do this, the ideas of using an information resource dictionary system (IRDS) like standardized in [ISO, 1990] are extended in a way, that data, meta data, and mapping of data can be stored and used to generate additional software in form of system add-ons or layers doing this mapping automatically.

In general, in order to meet the migration requirements of existing system components and legacy systems, standard data elements and their common representation must be accompanied by respective standard mapping rules. After having agreed on a common shared data model and the mapping rules for harmonization defined and distributed by the system independent data management organization, data mediation, i.e., the automatic translation of system dependent data elements into standardized data elements for information exchange and vice versa, becomes possible. The objective is to enable separate systems and system components, which have an overlap of interest, to interchange or share data in a common data representation.

The data mediation implementation described in this and respective further documents favors a software framework which may be linked as an additional software layer to existing systems and system components. It can be implemented as a common platform to migrate existing systems and system components and integrate future ones into an common shared data model-based interconnection network.

The data mediation approach is derived from database federation techniques, thereby, extending these techniques. In the concept having been introduced to the M&S/HLA[3] community by [Krusche and Tolk, 1999], data sources are no longer restricted to database systems. Any software component which produces and consumes data is considered as a „data storage medium". With this approach a data mediation framework is a common shell for any system component. The respective data mediation framework architecture summarizes these aspects in a common software platform architecture.

The data mediation approach enables any system component and any system with an individual data representation to be represented by an shared data model representation. This, however, requires to first harmonize individual data representations with the shared data model schema, which has to be done by the system independent data management agency described in the former section.

To summarize the ideas of this and the former section, the data harmonization necessary to gain a common understanding between two or more systems should be done in a way leading to standardized data elements. Information interchange is then done by using the respective SDEs. The harmonization process can be supported by tools resulting in mapping rules that can be used to program and configure drivers or layers enabling legacy systems to import and export the respective SDEs without having to reprogram the systems themselves. More detailed technical aspects are given in [Krusche and Tolk, 1999]. An awarded application of this techniques is described in [Tolk, 1999b] and [Tolk, 2000].

### 3.3 Technical Requirements for Simulation Systems

Much attention is paid to the use of simulation systems for Support to Operations, even if simulation systems only make up a small part of the possible OR methods which can be used as a decision support system according to the results of a German study conducted on behalf of the German Ministry of Defense [IABG, 1999]. Not least because corresponding simulation systems are well know to the officers from training and exercises (CAX - *Computer Assisted Exercise*), the requirement exists to use the known simulation systems not only for training purposes but also as decision support systems. In addition, support to operations as a domain for simulation systems can be increasingly found also in "political" papers like, e.g., [NATO, 1998].

However, recent studies pointed out that in order to enable simulation models to be used for this purpose, at least four core requirements have to be fulfilled by the respective system:

---

[3] HLA = High Level Architecture; IEEE P1516 (Draft) Standard for Simulation Federations.

1. The models must be verified, validated and accredited. In particular, all elements of the integrated system Command and Control– Reconnaissance – Effects must be realistically simulated, i.e., everything that may be relevant to a problem to be solved has to be modeled at least implicitly.
2. The data used in the models must be procurable. To the extent this is possible, the data are also to be verified, validated and accredited or certified. In addition, collecting, assessing and evaluating mass of data have to be supported.
3. The models must take the incremental increase of the situation into consideration and efficiently support its implementation into the simulated situation. As the real world data is vague, uncertain and contradictive in many cases, at least uncertainty management has to be integrated into simulation systems. In addition, new algorithms for creating the initial state of a simulation system out of a set of situation perceptions have to be created and validated, as this is a typical requirement only being needed in the domain application of decision support.
4. Command agents and alternative techniques from the Computer Generated Forces (CGF) sector must support the user with
   - the automatic compilation of orders,
   - the intelligent modeling of the enemy behavior and
   - the compilation of alternative courses of action in a given situation.

These requirements are at least partly well known from computer assisted exercises and other simulation application domains also, however, in the context of support to operations they have at least a new quality.

Support to operations means "Bringing Operations Research back to War". In an exercise, the missing of a model for C2, Reconnaissance, etc. will possibly lead to some exercise artifacts, but during an exercise it is easy to live with such things. During an operation, the missing of a respective component makes the use of the system hardly possible. In the same manner, in an exercise the data can be fitted to meet the purpose of the exercise executing group. During an operation, the data must be reliable. The use of command agents and computer generated forces is a must within operations, potentially facilitating exercises tremendously (and making them cheaper, as not so many personal has to be involved in the exercise). The common integration framework depicted also in the following figure can be reached by applying the findings of the last sections.

The last kernel requirement, however, is quite different. When doing analyses, simulation based acquisition and computer assisted exercises, a simulation system is used to evaluate the behavior of a dynamical system from a given state into the future. This is the case for all three domains of simulation systems: the simulation system starts from a well known situation. When using simulation systems for decision support, the data for the initial situation have to be extracted from the underlying C4I system first, i.e., the initial state of the simulation system must be created from the perceived situation being stored in the C4I system. As have been pointed out before, these are, however, vague, uncertain, incomplete, and contradictory data. In addition to this problem, these vague, uncertain, incomplete, and contradictory data have to be mapped to the sharp, certain, and complete data of the world of simulation systems. When doing operations support, the start from a given initial state is not the rule. One has to build this initial state from the perceived situation every time new information leading to new assumptions arrives. This, however, is a totally new type of requirement for simulation systems.

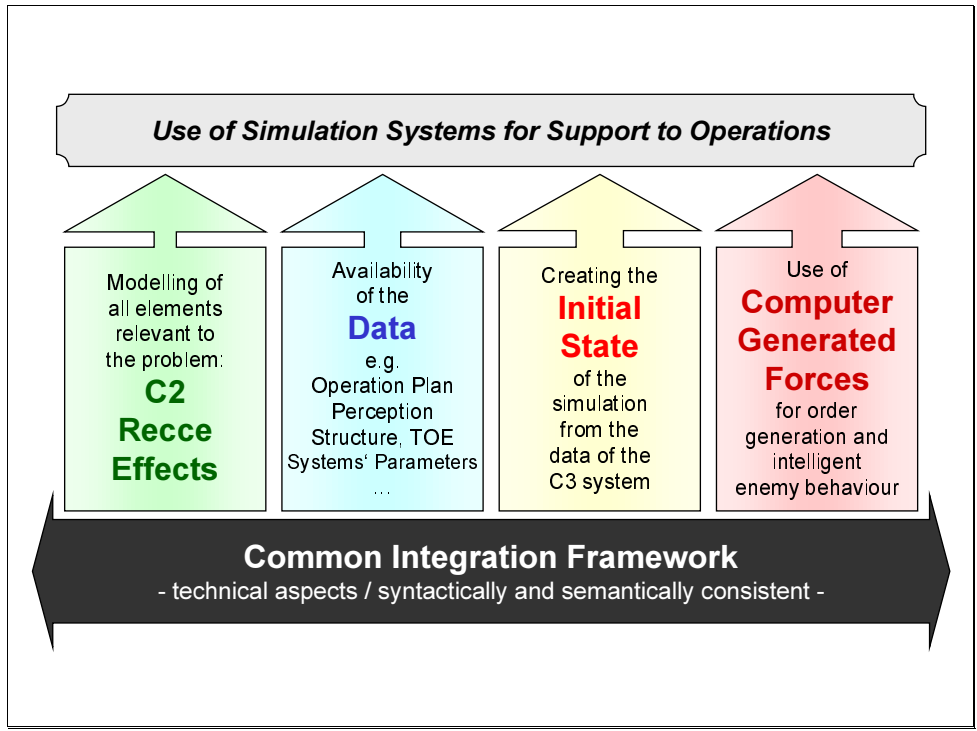

**Figure 4: Kernel Requirements for Simulation Systems**

It should be clear by this, that decision support is more than coupling C4I systems with legacy simulation systems. The requirements for simulation systems to be used as decision support systems varies from the requirements known from other application domains. More details can be found in the papers [IABG, 1999; Tolk, 1999a; Tolk and Schiefenbusch, 1999].

## 4       Operational Issues of Decision Support Systems

Decision support systems simplify and accelerate the information collection process, the provision of the same and its evaluation. Decision support systems do not serve the incorporation of automatisms. Therefore this concerns interactive systems. By use of the system, a military decision-maker obtains the support he needs in the form of information. Decision support systems do not form individual technical solutions.

For the assessment of the possibility of the use of decision support systems, however, not only the technical side of the matter is relevant. Rather important are the improvements to be expected to the whole Command and Control system by means of corresponding effects on the Command and Control process, Command and Control organization and means as a whole.Fundamentally what is at stake is to use decision support systems suitable in adequate terms of situation, task and level. The factors Time and Quality Improvement mark the significance of decision support systems for this particularly. However, the realized integrated system of Command and Control – Reconnaissance - Effects first enable the optimum use of decision support systems to be made.

The statements are thus first completely effective with the realization of this integrated system and the future development of corresponding OR procedures.

Decision support systems must aggregate or de-aggregate detailed structures for the enemy and for the own side depending upon the Command and Control level, task and situation.

However, it is obvious that all the decision support systems enable the link between a wide range of information, data and sectors to an integral "situation estimate" matching the situation, tasks and level to be made. With the further development of the Operations Research methods it will in future be increasingly possible to predict the behavior of complex systems or to assess alternative procedures and then in good time – based on the results of the analyses performed and parallel to that on-going analyses – in a control function to impact upon the processes really taking place.

Improvement of the decision-making preparation is the primary aim of all the behavior of complex systems in the planning phase. The development of alternative courses of action can be derived using decision support systems more logically and more rapidly than to date. Even if decision support systems themselves are not creative, they can support by means of the automatic compilation of alternatives. Decision support systems can in addition serve to elaborate risk sectors and unclarified problems in decisions. By means of decision support systems the commander can have his decision checked to the extent to which all boundary conditions are complied and all the relevant influence factors are taken into consideration. A forecast on the development of the combat is limited on the basis of the decision.

Effects on decision support systems on the operations plan are similarly envisaged in the optimization and in gaining time. In particular, the whole Command and Control process itself can be the aim of the optimization by means of decision support systems. The prerequisite for this however is a superior command post concept for the use of decision support systems in an integrated environment. By means of the continuously running forecast of the situation development using decision support systems, a short-term recontrol of 'running' operations is possible. Effects of the operation of decision support systems on the Command and Control organization are today at best recognizable as approaches. The effects will change repeatedly with the growing skills of decision support systems and are not the subject of the investigations of the study underlying this paper.

From the already presented basic thoughts on the modular use of decision support systems, it is considered necessary long-term to make decision support systems integrated into the operations system available, i.e. decision support systems must directly be usable at the workstation of the user. Insights from operations must continuously flow into the models.

The integrated decision support systems is long-term the actual aim of the further development process of the technical Command and Control means:
- the integrated system stores the information of the transparent battle space,
- the integrated system of decision support systems permits the suitable preparation of this information in terms of situation, task and level.

It must be possible to perform rapid data exchange between all existing or new information systems both for operations and for the peacetime staff tasks and with the various modules of decision support systems. From the migrated sub-systems and operational systems, the Integrated Information System for the Armed Forces is in the long term generated. Finally the integrated system must access, as the technical backbone of the transparent battle space considerably more information than systems to date.

From the analyzed effects of the use of decision support systems, demands are made of the military commander and his staff, which must be taken into consideration in education, training and exercises. The use of decision support systems makes high demands of the user. Decision support systems make continuous training of all members of military staff and the commanders at the C4I systems a must. This functions only if the same or similar systems are used in peacetime and for military operations. This means, that decision support systems and their use must be integrated as early a possible in training and daily use. What has to be ensured is that the user of decision support systems knows and understands the assumptions and limits of the system and the OR methods to be applied. Decision support systems training means therefore also training in OR methods.

The possibilities of prototype use by the military user of an evaluation model, i.e. the use of existing decision support systems, the inclusion of simulation systems and corresponding commercial software products, should be made full use of. In the medium term, the migration of existing systems to decision support systems (laboratory operation, interim solution, field usability) is to be targeted.

The compilation and documentation of a concept for the setting up, filling and maintenance of a basic database for decision support systems, taking the requirements made of the integrated information system of the Army or the Federal Armed Forces into consideration, are further pre-requisites for the successful use of decision support systems. Decision support systems are deployed in battle with information and for information superiority. They consist to this end of a wide range of individual modules which make their functionality available to the whole system via standardized interfaces. In particular automatable decisions and simple, pre-plannable routine decisions will determine – at least in the medium term – the combat areas of decision support systems. In this process the decision support systems support the military decision maker without replacing him.

## 5    Summary and the Way Ahead

As already having been said in the introduction, the merge of C4I components and OR components will characterize the future warfighter's information systems. The increasing use of common standards, data models, architecture, and reusable common components will facilitate the interoperability of functional components from both worlds leading on the long scale to heterogeneous federations comprising the functionality of both worlds. In the future, the distinction between systems for system modeling, acquisition, requirements testing, training and education, war fighting, online decision support (ACOA – alternative course of action development and analysis, DTTP – development of doctrine and tactics, techniques, and procedures), and AAR – after action review, and operation's evaluation will vanish more and more.

The deployment of decision support systems is already today technically feasible. Decision support systems will relieve the military commander of simple decisions and allow him more freedom action for his real Command and Control tasks and complex decisions. Moreover, they contribute directly to the fulfillment of the requirements for the Army of the new millennium, in that decision support systems make the operational requirements known, namely in that they support information superiority in operations and enable access to the information of the "transparent battle space" and its administration in own information responsibility, plan operations in the battle space, coordinate and help control, enable the need-oriented operations support to be given, support simultaneous operations, plan better protection of own forces, help coordinate and control and thus in the final instance contribute to qualitative superiority of own forces.

By use of decision support systems, in the long term it will be possible to cope with more tasks and information in a shorter period of time and to achieve better results and to bind less personnel by taking over routine tasks. They are kernel technologies techniques for advanced C4I systems supporting the commander in gaining the information superiority necessary in future operations scenarios.

*Acknowledgements*

*The authors like to thank Dr. Stefan Krusche for his contributions in the referenced articles. Many of the ideas have been designed by him for C4I system interoperability and have been adapted for new purposes.*

## 6 References

Following books and articles are referenced in the paper: